\documentclass[journal=acs,manuscript=article,layout=preprint]{achemso}
\newcommand{\onlinecite}[1]{[\hspace{-1 ex} \nocite{#1}\citenum{#1}]}
\usepackage[T1]{fontenc} 
\usepackage{color}
\usepackage[dvipdfmx]{}
\usepackage{ulem}

\author{Seiji Ueno}
\email{se-ueno@hpc.co.jp}

\affiliation[HPCS]{HPC Systems Inc., Japan}
\altaffiliation{Department of Chemistry, Kyoto University, Kyoto, Japan}

\author{Yoshitaka Tanimura}
\email{tanimura@kuchem.kyoto-u.ac.jp}
\altaffiliation{Department of Chemistry, Kyoto University, Kyoto, Japan}

\title[mytitle]
  {Modeling intermolecular and intramolecular modes of liquid water using multiple heat baths: Machine learning approach}
\abbreviations{}
\keywords{American Chemical Society, \LaTeX}

\begin{document}

\begin{abstract}
The vibrational motion of molecules in dissipative environments, such as solvation
 and protein molecules, is composed of contributions from both intermolecular and intramolecular modes.
 The existence of these collective modes introduces difficulty into quantum simulations of chemical and biological processes.
 In order to describe the complex molecular motion of the environment in a simple manner, we introduce a system-bath model
 in which the intramolecular modes with anharmonic mode-mode couplings are described by a system Hamiltonian,
 while the other degrees of freedom, arising from the environmental molecules, are described by heat bath.
 Employing a machine-learning based approach, we determine not only the system parameters of the intramolecular modes
  but also the spectral distribution of the system-bath coupling to describe the intermolecular modes,
  using the atomic trajectories obtained from molecular dynamics (MD) simulations.
 The capabilities of the present approach are demonstrated for liquid water using MD trajectories 
  calculated from the SPC/E model and the polarizable water model for intramolecular and intermolecular vibrational spectroscopies (POLI2VS)
  by determining the system parameters describing the symmetric-stretch, asymmetric-stretch and bend modes with intramolecular interactions
 and the bath spectral distribution functions for each intramolecular mode representing the interaction with the intra-molecular modes.
 From these results, we were able to elucidate the energy relaxation pathway between the intramolecular modes and the intermolecular modes in a non-intuitive manner
\end{abstract}

\section{1. INTRODUCTION}

Elucidating the effects of molecular environments on chemical and biological processes
 in both classical and quantum regimes has been an important problem in chemical physics over the last four decades.
 \cite{CaldeiraPhysica83,LeggettRMP87,Weiss12,Breuer02,TanimuraJPSJ06}
Theories of quantum open systems have been used to construct models of practical interest,
 \cite{Wolynes81,Miller1989,HanggiRMP90, BOmodel03,Onuchic85,Wolynes87ET, Ikeda18, Ikeda19, MasonHess89, SakuraiNJP14,Hanggi97,Hanggi09, KatoJPCB13, Levitt,Kampen81,Mukamel95,Khun95,TaniIshi09, Ishizaki09,Schuten12, Renger06, KramerAspu13, KramerAspu14, Coker2016}
 in particular to account for line shapes in NMR\cite{Levitt,Kampen81} and linear and nonlinear laser spectra.\cite{Mukamel95,Khun95,TaniIshi09}
The key feature of the environment is that it gives rise to irreversible dynamics through which the system evolves toward the thermal equilibrium state at finite temperature.\cite{TanimuraJPSJ06} 
The environmental effects that arise from solid state materials, solvation, and protein molecules are generally described
 by an interaction between a primary molecular system coupled to a harmonic oscillator bath (HOB).
Two commonly used models of this kind are the Caldeira-Leggett (Brownian) model\cite{LeggettRMP87}, 
Brownian Oscillator model (BOM)
and the spin-Boson model.\cite{LeggettRMP87} 
The HOB model, whose distribution takes a Gaussian form, exhibits wide applicability, despite its simplicity.
This is because the influence of the environment can in many cases be approximated by a Gaussian process,
 due to the cumulative effect of the large number of weak environmental interactions.
In such a situation, the ordinary central limit theorem is applicable, and hence the Gaussian distribution function is appropriate.\cite{TanimuraJPSJ06}
The distinctive features of the HOB model are determined by the spectral distribution function (SDF) of the coupling strength, $J(\omega)=\sum c_j^2 \delta(\omega-\omega_j)$,
 where $c_j$ is the coupling strength between the system and the $j$th bath oscillator, with frequency $\omega_j$. 
By properly choosing the form of the SDF, the properties of the bath can be adjusted
 to represent a variety of environments consisting of, for example, solvates and protein molecules. 

Several analytical and numerical approaches have been developed for such models to treat the quantum and classical motion of molecules in condensed phases.
These include the generating functional approach\cite{Tanimura93,OkumuraPRE96,TanimuraOkumuraJCP97},
 the Redfield equation approach,\cite{ModifedRedfield06,Domcke08,Reichman15}
 the quasi-adiabatic propagator path integral (QUAPI) approach, \cite{Makri96,Makri2014}
 the multi-configurational time-dependent Hartree (MCTDH) approach, \cite{ML-MCTDH1,ML-MCTDH2,WangTHoss1,ML-MCTTDHRev}
 and the reduced hierarchy equations of motion (HEOM) approach.\cite{Tanimura89A, TanimuraPRA90,IshizakiJPSJ05, Tanimura2014,Tanimura2015,Xu05,Yan06, Wu15, Ulrich16,Tanimura18}
These approaches have been applied to problems of practical interest,
 in particular to the investigation of chemical reaction processes, \cite{Wolynes81,Miller1989,HanggiRMP90, BOmodel03} 
 non-adiabatic transitions,\cite{Onuchic85,Wolynes87ET, Ikeda18, Ikeda19} 
 quantum device systems,\cite{MasonHess89, SakuraiNJP14} 
 ratchet rectification,\cite{Hanggi97,Hanggi09, KatoJPCB13}
 exciton transfer,\cite{Ishizaki09,Schuten12, Renger06, KramerAspu13, KramerAspu14, Coker2016} 
 and the analysis of the linear and nonlinear laser spectra.\cite{Mukamel95,Khun95,TaniIshi09}

Because the key feature of environments is determined by the system-bath coupling and the SDF, a methodology to determine these is significant. 
Typically, the SDF is estimated from linear and nonlinear infrared and Raman spectra,
 both experimentally\cite{IR_exp1,Palese96,Raman, Raman1_exp,Raman_exp2, Pullerits20}
 and numerically.\cite{Hasegawa2DRaman,2DTR-TM01,2DIRr,H-Water, JJLiu18,Xanthes2008, Renger06, KramerAspu13, KramerAspu14} 
The SDFs of entire molecular systems have also been evaluated directly from molecular dynamics (MD) trajectories
 on the basis of normal modes\cite{ChoFlemingOhmineStratt94,INM01,INM02,ImotoSaito2013A, ImotoSaito2013B,Yagasaki2011, ImotoSaito2015, 2DIR-MD4,2DIR-MD5,Okazaki1998, Okazaki2000,Okazaki2003},
 the velocity-velocity autocorrelation function,\cite{INM01,Yagasaki2011} and Fourier transformations.\cite{Kuehn2019}
While the method based on optical spectra does not allow estimation of optically inactive system modes, the approaches based on MD trajectories have the capability to describe all intermolecular and intramolecular vibrational motions.
However, in such trajectory-based methods, it is not easy to separate the system and the heat-bath, in particular in the case of the intermolecular modes.
This is because these methods are based on naive statistics of the complex and irregular motion of molecular trajectories, and thus, with these methods, it is not easy to distinguish the bath modes from the background noise.
Moreover, because the SDF generally depends on the system parameters, for example depending on the intermolecular mode-mode coupling strengths, with the existing methods, it is not easy to obtain the system parameters and the bath SDF simultaneously.

In this study, we employ a data mining approach, which allows us to obtain variable information from massive data in physics and chemistry\cite{Datamining1,Datamining2,Datamining3} on the basis of the statistical analysis.\cite{DataminingBook, BishopPRML} 
We then develop an efficient algorithm to optimize both the system parameters and the SDF through analysis of MD trajectories.
This algorithm allows us to obtain not only the SDF of each intramolecular mode, but also the variance of the bath-correlation function that describes the vibrational dephasing of molecular modes.
We demonstrate this approach for liquid water using MD trajectories obtained from SPC/E model\cite{gromacs1,gromacs2} and POLI2VS model.\cite{H-Water,JJLiu18} 

This paper is organized as follows. In Sec. ~2, we introduce a multi-mode system-bath model and describe the data-mining approach that we use to determine the system parameters, the system-bath interaction, and the SDFs. 
In Sec. 3, we present model potentials for intramolecular modes of liquid water in the framework of BOM.
In Sec. 4, we present the details of the calculation.
In Sec.5, we present the results for liquid water obtained through analysis of the MD trajectories obtained using various force fields. Section 6 is devoted to concluding remarks.

\section{2. THEORY}
\label{sec:Theory}
\subsection{A model with multiple heat baths}

In order to describe the vibrational modes of molecular liquids, we consider a model that consists of primary intramolecular modes coupled to intermolecular environmental modes, which are regarded as bath systems.
These bath systems are represented by ensembles of harmonic oscillators. \cite{Mukamel95}
The model is constructed by extending a Brownian (or Caldeira-Leggett) Hamiltonian
 to include a nonlinear system-bath interaction,\cite{Oxtoby1976, Oxtoby1979, Berne1994,Okumura-PRE-1997-56, dephase04,dephase05,dephase06,dephase07, Sakurai-JPCA-2011-115}
 which causes the frequency and amplitude of the intramolecular modes to vary in time or to be inhomogeneously distributed.
We can describe both situations within a unified framework by adjusting the nonlinear system-bath coupling strength.\cite{TanimuraJPSJ06,TaniIshi09} 
The total Hamiltonian is expressed as
\begin{eqnarray}
H=\sum_s\left(H^{(s)}_{\mathrm{S}}+H^{(s)}_{\mathrm{B}}+H^{(s)}_{\mathrm{I}} \right) + \sum_{s \ne s'}U_{s,s'}(q_s, q_{s'}),
  \label{eq:h_total}
\end{eqnarray}
where
\begin{eqnarray}
  H^{(s)}_\mathrm{S}=\frac{p_s^2}{2m_s}+U_s(q_s)
  \label{eq:h_system}
\end{eqnarray}
is the Hamiltonian for the $s$th mode, with mass $m_s$, coordinate ${q_s}$, momentum ${p_s}$, and potential $U_s(q_s)$.
The interaction between the modes $s$ and $s'$ is given by 
\begin{equation}
U_{s,s'}(q_s, q_{s'}) = g_{11}q_sq_{s'} + \frac{1}{2}\left(g_{21}q_s^2q_{s'} + g_{12}q_sq^2_{s'}\right).
  \label{eq:U_couple}
\end{equation}
The forms of the potentials and interactions should be chosen to accurately describe the system dynamics in a simple manner.
The bath Hamiltonian for the $s$th mode is expressed as
\begin{eqnarray}
H^{(s)}_\mathrm{B}=\sum _{j_s}\left(\frac{p_{j_s}^2}{2m_{j_s}}+\frac{m_{j_s}\omega _{j_s}^2x_{j_s}^2}{2} \right)
  + \sum _{j_s}\left(\frac{\alpha_{j_s}^2V_s^2({q_s})}{2m_{j_s}\omega _{j_s}^2}\right),
 \label{eq:h_bath}
\end{eqnarray}
where the momentum, coordinate, mass, and
frequency of the $j_s$th bath oscillator are given by $p_{j_s}$, $x_{j_s}$, $m_{j_s}$ and $\omega_{{j_s}}$, respectively.
The last term in the above is a counter term, which maintains the translational symmetry of the system in the case ${U}_s(q_s) = {U}_{s,s'}(q_s, q_{s'})=0$.\cite{CaldeiraPhysica83} 
The system-bath interaction is given by
\begin{equation}
  H^{(s)}_{\mathrm{I}} =- V_s(q_s)\sum_{j_s}\alpha_{j_s}x_{j_s},
  \label{eq:h_int}
\end{equation}
which consists of linear-linear (LL) and square-linear (SL) interactions,
\begin{equation}
V_s(q_s)\equiv V^{(s)}_{\mathrm{LL}}q_s+\frac{1}{2} V^{(s)}_{\mathrm{SL}}q_s^2,
\label{eq:V_interact}
\end{equation}
 with coupling strengths $V^{(s)}_{\mathrm{LL}}$, $V^{(s)}_{\mathrm{SL}}$, and $\alpha_{j_s}$.\cite{TanimuraJPSJ06}
As shown in Refs. \onlinecite{TanimuraJPSJ06,TaniIshi09}, while the LL interaction contributes mainly to energy relaxation,
 the SL system-bath interaction leads to vibrational dephasing in the slow modulation case,
 due to the frequency fluctuation of the system vibrations.\cite{ dephase04,dephase05, dephase06,dephase07, Ishizaki-JCP-2006-125,Sakurai-JPCA-2011-115} 
Then, combining Eq.(\ref{eq:h_bath}) and Eq.(\ref{eq:h_int}), we obtain 
\begin{equation}
H^{(s)}_\mathrm{B} + H^{(s)}_\mathrm{I} =
 \sum _{j_s}\left(\frac{p_{j_s}^2}{2m_{j_s}}+\frac{m_{j_s}\omega _{j_s}^2\tilde{x}_{j_s}^2}{2} \right),
\label{eq:h_bathint}
\end{equation}
where $\tilde{x}_{j_s} = x_{j_s} - ({\alpha_{j_s}V_s({q_s})})/({2m_{j_s}\omega^2_{j_s}})$ is the re-oriented bath coordinate.
This model has been used to derive predictions for the study of single-mode systems
 employing 2D Raman\cite{dephase04,dephase05,dephase06,dephase07}, 2D THz-Raman,\cite{2DTR-TM01} and 2D IR signals,\cite{Ishizaki-JCP-2006-125,Sakurai-JPCA-2011-115}
 and for the study of two-mode systems employing 2D IR\cite{2DIR-T01,2DIR-T02} and 2D THz-IR signals.\cite{2DIRr} 
Here, we apply this model to describe the intermolecular modes of molecular liquids. 
We assume that the influences of the fluctuation and dissipation on individual modes are all independent
 and that the correlations of the fluctuations among different modes can be ignored.\cite{2DIR-T01,2DIR-T02}
Then, the SDF is defined as
\begin{equation}
J_s(\omega )\equiv \hbar \sum _{j_s}\frac{\alpha _{j_s}^2}{2m_{j_s}\omega_{j_s}} \delta (\omega -\omega_{j_s}),
\label{eq:def_Js}
\end{equation}
This definition characterizes the nature of the bath. 

\subsection{Optimizing the likelihood probability distribution}

Let us onsider the trajectory of the $k$th molecule in phase space, expressed as $\left(\mathbf{q}_k(t), \mathbf{p}_k(t)\right)$, 
 describing the intramolecular motion of interest obtained from the MD simulation.
 We compute the set of trajectories $\left(\mathbf{q}_k(t_i+ i\Delta t), \mathbf{p}_k(t_i+i\Delta t)\right)$
  with time step $\Delta t$ for all integer $i$ satisfying $0\le i \le N-1$,
 where $N$ is the total number of time steps. Using a data mining approach,
  we attempt to reproduce the MD trajectories for the intramolecular modes
 by adjusting the system parameters in Eq.(\ref{eq:h_total}) and the SDF for each system mode.
 The trajectory of the $j_s$th bath oscillator for the $s$th system mode in Eqs. (\ref{eq:h_bath}) and (\ref{eq:h_int}) is assumed to take the form
\begin{equation}
\tilde{x}_{j_s}(t) = A_{j_s} \sin(\omega_{j_s} t + \phi_{j_s}),
\label{Xjs}
\end{equation}
where $A_{j_s}$ and $\phi_{j_s}$ are the amplitude and phase of the ${j_s}$th bath oscillator.
The phase $\phi_{j_s}$ is chosen randomly to avoid recursive motion of the oscillator.
This implies that the bath oscillators described by Eq.(\ref{eq:h_bathint}) are harmonic for any form of the system-bath coupling.

In the LL coupling case, the bath parameters and the system-bath interactions are expressed as a set of latent variables in the machine learning context, defined as 
\begin{equation} 
\mathbf{z}_k = \left(
  \left\{c^k_{j_1} \right\},
  \left\{c^k_{j_2} \right\},
  \cdots,
  \left\{c^k_{j_N} \right\}
\right)
\label{eq:def_zk}
\end{equation}
where $\left\{c^k_{j_{s}} \right\}$ is the set of bath coupling parameters with the element, 
\begin{equation}
c^k_{j_s} = V_\mathrm{LL}^{(s)}\alpha^k_{j_s}A^k_{j_s},
\label{eq:def_ck}
\end{equation}
for the mode $s$.

A set consisting of the system potential parameters and the SDFs is denoted by $\mathbf{\Sigma}$. 
In the present approach, these are optimized in the same manner.
The trajectory at time $t_i+ i \Delta t$ for integer $i$ ($1\le i \le N$) that is calculated using the HOB model on the basis of the MD trajectories at $t_i+(i-1)\Delta t $  is then expressed as
\begin{equation}
 (\tilde\mathbf{q}_k(t_i+ i \Delta t), \tilde\mathbf{p}_k(t_i+ i \Delta t))
  = \hat{L}(\Delta t; \mathbf{z}_k, \mathbf{\Sigma})( \mathbf{q}_k(t_i+(i-1)\Delta t), \mathbf{p}_k(t_i+(i-1)\Delta t)),
\end{equation}
where $\hat{L}(\Delta t; \mathbf{z}_k, \mathbf{\Sigma})$ is the Liouville operator of $\mathbf{q}_k $ and $\mathbf{p}_k $
 for the model Hamiltonian (\ref{eq:h_total}) with the bath functions given in Eq. (\ref{Xjs}).
The differences between the coordinates and momenta for a given MD trajectory and the corresponding HOB trajectory are defined as 
$\Delta \mathbf{q}_k (t_i+ i \Delta t)\equiv \left(\tilde{\mathbf{q}}_k(t_i+ i \Delta t) - \mathbf{q}_k(t_i+ i \Delta t)\right)$ and 
$\Delta \mathbf{p}_k (t_i+ i \Delta t)\equiv \left(\tilde{\mathbf{p}}_k(t_i+ i \Delta t) - \mathbf{p}_k(t_i+ i \Delta t)\right)$.

It should be noted that the influence of the environmental molecules can in many cases be approximated by a Gaussian process,
 because it can be treated as the cumulative effect of a large number of weak environmental interactions, in which case the ordinary central limit theorem is applicable.
In this case, the distribution function of the HOB model, which is Gaussian, is appropriate.
Thus, the difference between the actual influence of the environmental molecules
 and that described by the HOB model is due to the non-Gaussian aspect of the molecular motion.

Moreover, if we choose the time step, $\Delta t$, to be sufficiently large in comparison with the time scale of the frequency fluctuations of the intramolecular modes, 
 we can regard the dynamics to be Markovian, in which case $\Delta \mathbf{q}_k (t_i+ i \Delta t)$ and $\Delta \mathbf{p}_k (t_i+ i \Delta t)$
 for different $i$ are not correlated. Here, we indeed consider this situation.
Thus, we assume that the time evolution of these variables obeys a Gaussian-Markovian process,
 while the intramolecular modes themselves are described by the system Hamiltonian. 

The mean squares of the differences given above are is expressed as
\begin{equation}
  \sigma_{qq} \delta \mathbf{q}_k^2(\Delta t) = \frac{1}{N}\sum_{i=1}^{N} \Delta \mathbf{q}_k^2 (t_i+i\Delta t)
  \label{eq:def_sigmaqq}
\end{equation}
and
\begin{equation}
  \sigma_{pp} \delta \mathbf{p}_k^2(\Delta t) = \frac{1}{N}\sum_{i=1}^{N} \Delta \mathbf{p}_k^2 (t_i+i\Delta t),
  \label{eq:def_sigmapp}
\end{equation}
where $\sigma_{qq}$ and $\sigma_{pp}$ represent the deviations between the BOM and MD trajectories.
 We next introduce a joint probability distribution function
  $ \delta P\left(\mathbf{q}_k(\Delta t), \mathbf{p}_k(\Delta t)\mid \mathbf{q}_k, \mathbf{p}_k; \mathbf{z}_k, \mathbf{\Sigma} \right)$,
 which describes the time evolution of the probability distribution
  from the state at time $t$ to that at time $ t + \Delta t $ 
  for a given set of system parameters, $\mathbf{\Sigma}$, and bath parameters, $\mathbf{z}_k$:
\begin{equation}
 \delta P\left( \mathbf{q}_k(\Delta t), \mathbf{p}_k(\Delta t)\mid \mathbf{q}_k, \mathbf{p}_k; \mathbf{z}_k, \mathbf{\Sigma} \right) =
 \exp \left[ - \sigma_{qq} \delta \mathbf{q}_k^2 (\Delta t) - \sigma_{pp}\delta \mathbf{p}_k ^2 (\Delta t) \right].
  \label{eq:prob_timestep}
\end{equation}

The parameters $\sigma_{qq} $ and $ \sigma_{pp}$ indicate the accuracy of the BOM.
In ordinary situations, these take finite values, because the description of molecular trajectories based on the simple BOM is limited,
 and because the choice of the parameters $\mathbf{z}_k$ may not yield the best fit.
We determine these parameters using the maximum likelihood series estimating method (MLSEM) and the gradient estimating method.
To use MLSEM, we maximize the logarithm likelihood ratio, defined by
\begin{equation}
  G(\mathbf{z}_k) = \log\left[\delta P\left(
         \mathbf{q}_k(\Delta t), \mathbf{p}_k(\Delta t)
    \mid \mathbf{q}_k, \mathbf{p}_k; \mathbf{z}_k, \mathbf{\Sigma} \right)\right].
  \label{eq:def_G}
\end{equation}
Then, using Eqs. (\ref{eq:def_sigmaqq})-(\ref{eq:def_G}), we obtain
\begin{equation}
    G(\mathbf{z}_k) = \frac{1}{N}\sum_{i=1}^N\left[\Delta\mathbf{q}^2_k(t_i + i\Delta t) + \Delta\mathbf{p}^2_k(t_i + i\Delta t) \right],
  \label{eq:def_G_2}
\end{equation}
which can be interpreted as the mean squared error of the BOM trajectories from the MD trajectories.

We then estimate $\mathbf{z}_k$ using the gradient estimating method for a given trajectory for the $k$th molecule,
\begin{equation}
\mathbf{z}_k = \mathop{\rm arg~min}\limits_{\mathbf{z}_k}G(\mathbf{z}_k).
\end{equation}
We repeat this procedure for all system molecules, and compute the average over these molecules in order to obtain the optimized set of $\mathbf{z}$.

\subsection{The bath spectral distribution functions}
\label{sec:SDF}
The total energy of the $j_s$ th oscillator is given by
\begin{equation}
  E_{j_s} = m_{j_s}\omega_{j_s}^2 (A^k_{j_s})^2.
\end{equation}
We assume that the probability distribution for each harmonic oscillator is described by the canonical ensemble with energy $E_{j_s}$.
The expectation values of the amplitude $A^k_{j_s}$ are then evaluated as
\begin{equation}
  \left<A^k_{j_s}\right> = \frac{1}{\sqrt{\pi\beta m_{j_s}\omega_{j_s}^2}},
 \label{eq:avg_Ak}
\end{equation}
where $\beta = 1/k_\mathrm{B}T$ is the inverse temperature of the system.
Substituting $\alpha^k_{j_s}$ into Eq. (\ref{eq:def_Js}), and using Eqs. (\ref{eq:def_ck}), and using (\ref{eq:avg_Ak}), we obtain
\begin{equation}
J^k(\omega_{j_s}) \sim \frac{\pi^2}{2V_\mathrm{LL}^{(s)}}\beta\omega_{j_s}(c^k_{j_s})^2.
\end{equation}

The SDF for the $s$th mode is evaluated as the average of $J^k(\omega_{j_s})$ over the samples $k$.

\section{3. INTRAMOLECULAR SYSTEM MODEL FOR LIQUID WATER}
\label{sec:SBmodel}
Here, we demonstrate the data-mining approach for liquid water to construct the system-bath Hamiltonian.
The system Hamiltonian for the intramolecular vibrational modes of a water molecule consists of the symmetric-stretch, asymmetric-stretch and bending modes,
 which we assume to take the form
\begin{equation}
 \sum_s H^{(s)}_{\mathrm{S}} = \sum_{\alpha\in\{\mathrm{O, H_1, H_2}\}}\left(\frac{1}{2} m_\alpha\vec{v}_\alpha^2\right) + U(r_1, r_2, \theta),
\end{equation}
where $m_\alpha$ and $\vec{v}_\alpha$ represent the mass and velocity of the O, H1 and H2 atoms, and $U(r_1, r_2, \theta)$ is the water model potential.

We employ two models, a harmonic oscillator (HO) based model, defined as 
\begin{equation}
 U_\mathrm{HO}(r_1, r_2, \theta) = k_r\left(r_1 - r_0\right)^2 + k_r\left(r_2 - r_0\right)^2 + k_t\left(\theta_1 - \theta_0\right)^2,
\label{eq:BO}
\end{equation}
where $k_r$ and $k_t$ are the force constants for the OH stretching and HOH bending motion, and a Morse oscillator (MO) based model, defined as
\begin{eqnarray}
 U_\mathrm{MO}(r_1, r_2, \theta) &&= D\left(e^{-2a(r_1 - r_0)} - 2e^{-a(r_1 - r_0)}\right) \nonumber \\
&&+ D\left(e^{-2a(r_2 - r_0)} - 2e^{-a(r_2 - r_0)}\right) + k_t\left(\theta_1 - \theta_0\right)^2,
\label{eq:MO}
\end{eqnarray}
where $D$ is the dissociation energy, and $a$ is the width parameter.
In the present study, the other intermolecular vibrational modes are treated as the bath oscillators, described by $H_\mathrm{I}^{(s)}$ and $H_\mathrm{B}^{(s)}$.\cite{dephase04,dephase05,dephase06,dephase07,2DTR-TM01,Ishizaki-JCP-2006-125,Sakurai-JPCA-2011-115,2DIR-T01,2DIR-T02,2DIRr}
When the intramolecular and intermolecular modes are not well separated, the profiles of SDFs become very sensitive for a choice of system potential.
We found that the optimized SDFs for the HO and MO models differ significantly in the POLI2VS case, while they are similar in the SPC/E flexible case.
For this reason, here we employ the HO model in the SPC/E flexible case and the MO model in the POLI2VS case, setting $D = 444.188$ kJ/mol.
The equilibrium values of $r_0$ and $\theta_0$ for these models are evaluated through use of $\left<r\right>$ and $\left<\theta\right>$.
The trajectories obtained with this procedure are described in the $xyz$ coordinates for each atom.
The loss function given in Eq. (\ref{eq:def_G}) is evaluated by comparing the predicted $xyz$ trajectories with the MD $xyz$ trajectories.

We describe the intra-vibrational modes in terms of the two HO bond lengths and the HOH bond angle of a water molecule as
\begin{eqnarray}
  &&r_1= \left|\vec{x}_\mathrm{O} - \vec{x}_\mathrm{H1}\right|, \label{eq:def_molcoord1}\\
  &&r_2 = \left|\vec{x}_\mathrm{O} - \vec{x}_\mathrm{H2}\right|, \label{eq:def_molcoord2}\\
  &&\theta= \arccos \left( \frac{\left(\vec{x}_\mathrm{O} - \vec{x}_\mathrm{H1}\right)
  \cdot\left(\vec{x}_\mathrm{O} - \vec{x}_\mathrm{H2}\right)}{r_1r_2}\right),
  \label{eq:def_molcoord3}
\end{eqnarray}
where $\vec{x}_\mathrm{O}$ and $\vec{x}_{\mathrm{H}_k}$ are the positions of the oxygen atom and the $k$th hydrogen atom.

The system coordinates for the symmetric-stretch, asymmetric-stretch, and bend modes are expressed as
\begin{eqnarray}
&&q_\mathrm{sym} = \frac{1}{2}\left(r_1 + r_2 - 2r_0\right), \label{eq:def_modecoord1}\\
&&q_\mathrm{asym} = \frac{1}{2}\left(r_1 - r_2\right), \label{eq:def_modecoord2}\\
&&q_\mathrm{bend} = \theta - \theta_0,
  \label{eq:def_modecoord3}
\end{eqnarray}
where $r_0$ is the equilibrium length of the OH bond.
The system and system-bath interaction, $H^{(s)}_\mathrm{B}$ and $H^{(s)}_\mathrm{I}$, are described with these system coordinates.
The intramolecular mode-mode interaction, $U_{s,s'}(q_s, q_{s'})$, in the system Hamiltonian can be obtained by rewriting the intramolecular potentials
 appearing in Eqs.(\ref{eq:BO}) and (\ref{eq:MO}) in terms of Eqs.(\ref{eq:def_modecoord1})-(\ref{eq:def_modecoord3}), respectively. 
In the present approach, the system parameters, such as $k_r$ and $k_t$, may vary during optimization.
For this reason, we also optimize the mode-mode coupling constants $g_{ij}$ given in Eq. (\ref{eq:U_couple}).

\section{4. CALCULATION DETAILS}
\label{sec:Calculation}

The MD trajectories were obtained by carrying out the simulations for liquid water with 216 molecules in a cubic box with periodic boundary conditions.
We employed GROMACS 2018 for the flexible SPC/E model\cite{gromacs1,gromacs2} and the code developed by Hasegawa for the POLI2VS model.\cite{H-Water}
The equations of motion were integrated using the velocity-Verlet algorithm with ${\Delta}t = 0.1$ fs.
The MD trajectories were constructed for each 0.2 fs up to 300 fs at the temperature about $300$ K.

Because we model the fast intermolecular modes, which arise from short-range intermolecular interactions,
 it is not necessary to carry out large-scale simulations with many molecules.\cite{jyon16}
The external motion of molecules, i.e., the translational and rotational motion, was then eliminated from the obtained trajectories.
The time step $\Delta t$ for the BOM Hamiltonian, given in Eq.(\ref{eq:h_total}), was set to 0.1 fs.
The initial phase, $\phi_{j_s}^k$, was generated from a random number uniformly distributed over $[0, \pi)$ at every iteration step.
We chose the number of bath oscillator for the $s$th mode as $N_s = 1500$.
The frequency of oscillators $\omega_{j_s}$ was set to $j_s\Delta\omega$ with $\Delta\omega = 1\ \mathrm{ps}^{-1} \approx 5.31\ \mathrm{cm}^{-1}$; while almost all features of spectroscopic signals of water can be described within this resolution, the highest frequency of bath modes is $8000$ cm$^{-1}$ that covers the overtone of OH stretching mode.

We developed the Python code using the TensorFlow library\cite{tensorflow} to employ the Adam algorithm\cite{AdamOpt} for minimization of the loss function Eq. (\ref{eq:def_G}).
The learning rate of the Adam algorithm was set to 0.01.
The training data were obtained from the classical MD simulations: the 3888 trajectories for POLI2VS and the 216 trajectories for SPC/E were provided, respectively.
The initial values for $\mathbf{z}_k$ was set to 0.
The optimizations for the system parameters and $\mathbf{z}_k$ were then carried out. We chose 22 and 108 trajectories as a minibath for the SPC/E flexible case and the POLI2VS case, respectively. The same trajectories were used for each optimization epoch. The iterations were repeated until the optimization parameters of the potentials converged.
\begin{table}[htbp]
  \caption{Equilibrium OH bond length and HOH angle for SPC/E flexible and POLI2VS force field models.}
  \label{table:equilibrium_len}
  \begin{tabular}{ccc}
    \hline
    Model          & $r_0$      & $\theta_0$ \\ \hline \hline
    SPC/E Flexible & 0.10282 nm & 1.8177 rad \\
    POLI2VS        & 0.09779 nm & 1.8379 rad \\ \hline
  \end{tabular}
\end{table}

\section{5. RESULTS AND DISCUSSION}
\label{sec:Results}

\begin{figure}[htbp]
\includegraphics[bb=0 0 403 302, width=8cm]{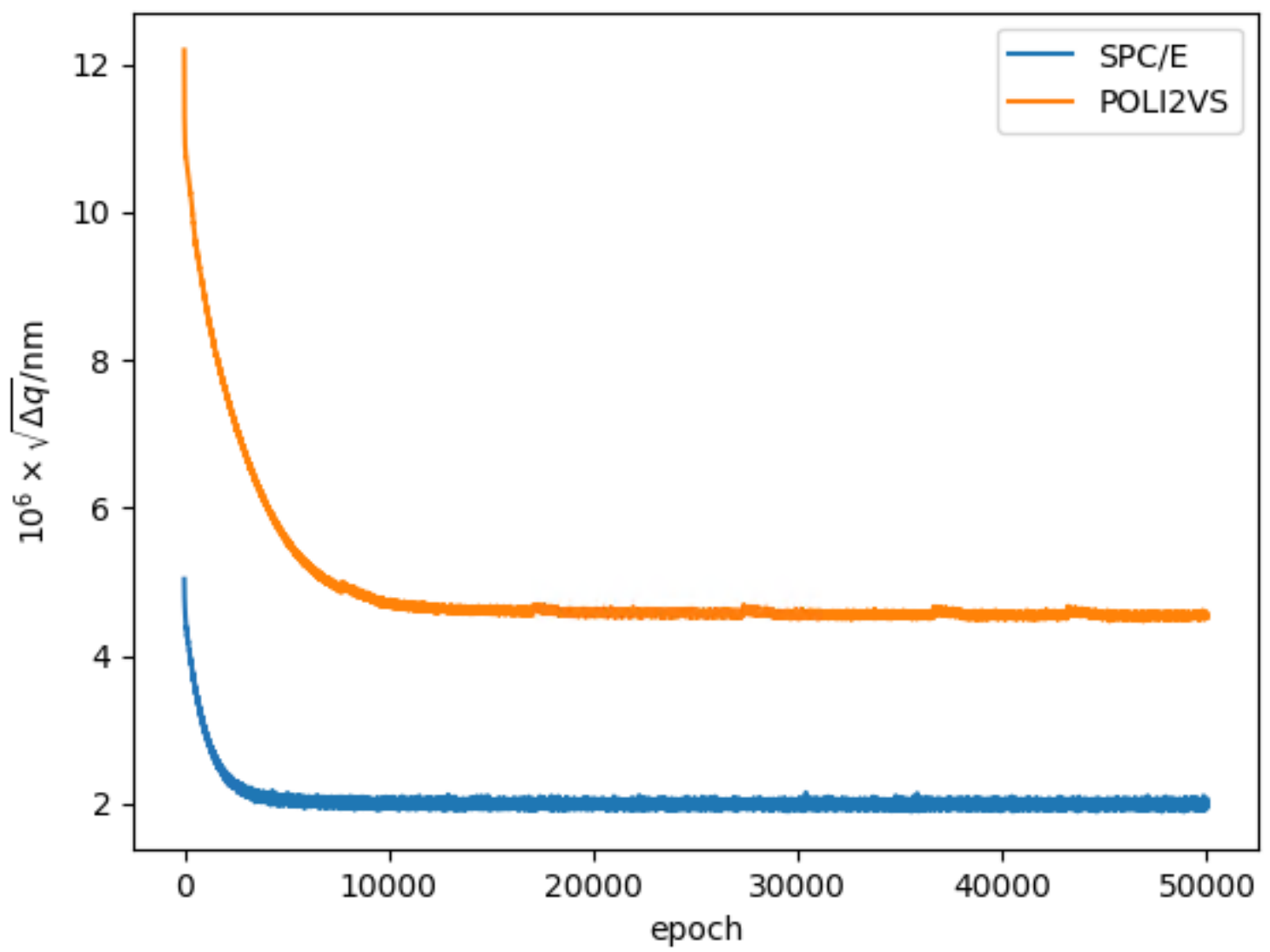}
\caption{
The learning curve of the mean squared error for $q$ between the predicted trajectories and the actual MD trajectories for the SPC/E flexible (blue curve) and the POLI2VS force fields (orange curve).}
\label{fig:pos_mse}
\end{figure}

\begin{figure}[htbp]
\includegraphics[bb=0 0 407 306, width=8cm]{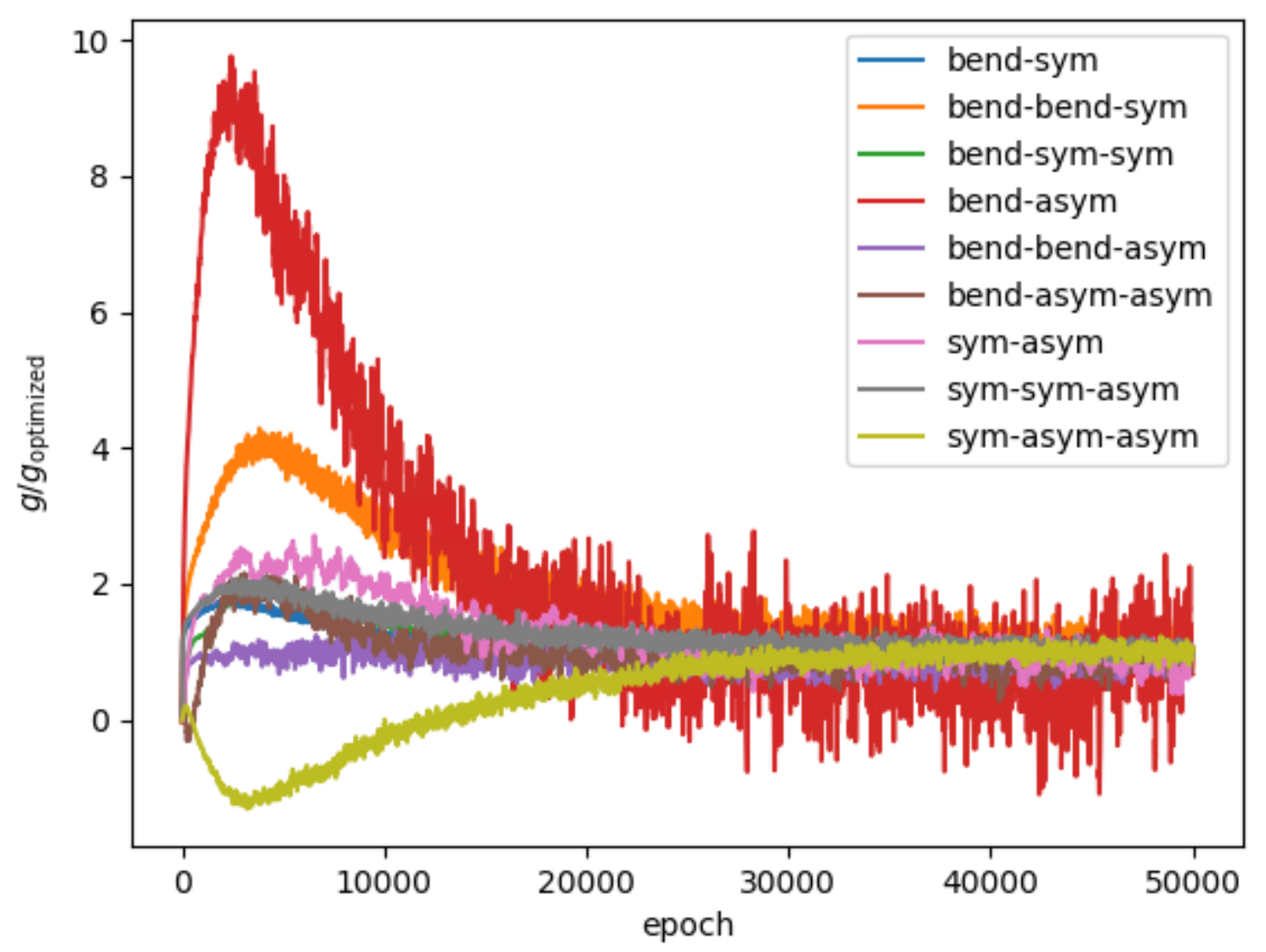}
\caption{
The learning curve of the intramolecular mode-mode coupling strength for the trajectories of the SPC/E flexible model. The bend-sym curve, for example, represents $g_{11}$ for the bending-symmetric coupling, while the bend-bend-sym and bend-sym-sym curves, for example, represent $g_{21}$ and $g_{12}$ for the bending-symmetric coupling, respectively.}
\label{fig:gij_curve}
\end{figure}

The optimized parameters for the potential, the intramolecular mode-mode coupling, and the linear and nonlinear system-bath couplings are listed in Table \ref{table:optimized_param}.
We employed approximately 50000 epochs for the optimization of the potential parameters, while we needed around 10000 epochs for the optimization of $J_{s}(\omega)$.
 We display the learning curves for the mean square of the coordinate, $\Delta q$, and the mode-mode coupling parameters, $g_{ij}$, in Figs. \ref{fig:pos_mse} and \ref{fig:gij_curve}.
The frequencies of the modes determined from the force constants and the parameters in the HO and MO models
 are found to be smaller than the values determined from the force field model used in the MD simulations.
This is because the LL and SL interactions enhance the system frequency,
 as shown by analytical and numerical analyses in the harmonic case.\cite{Okumura-PRE-1997-56, dephase04,dephase05,dephase06,dephase07,Sakurai-JPCA-2011-115}
The anharmonic mode-mode coupling constants differ significantly in the HO and MO cases.
This is because some of the contributions to the anharmonic mode-mode coupling have already been taken into account by the anhrmonicity of the MO potential, as explained in Sec. 3.
The obtained $V_\mathrm{SL}/V_\mathrm{LL}$ values indicate that the SL interaction plays a larger role in the symmetric-stretch case than in the asymmetric-stretch case,
 while the effect of the SL interaction is negligibly small in the bend case.
This is due to the formation of the hydrogen bonding (HB) between the system molecule and the surrounding molecules,
 which changes the vibrational frequencies of the O-H1 and O-H2 bonds.
The formation of the HB bonds for both the O-H1 and O-H2 bonds occurs in a correlated manner in the symmetric case,
 while it occurs in an anti-correlated manner in the asymmetric case.
Because the two formations occur simultaneously in the correlated case, the symmetric-stretch mode exhibits a larger SL effect.
Contrastingly, because the O-H bond lengths do not change significantly in the bending case, the effect of SL coupling is very small.

\begin{figure}[htbp]
\includegraphics[bb=0 0 360 576, width=8cm]{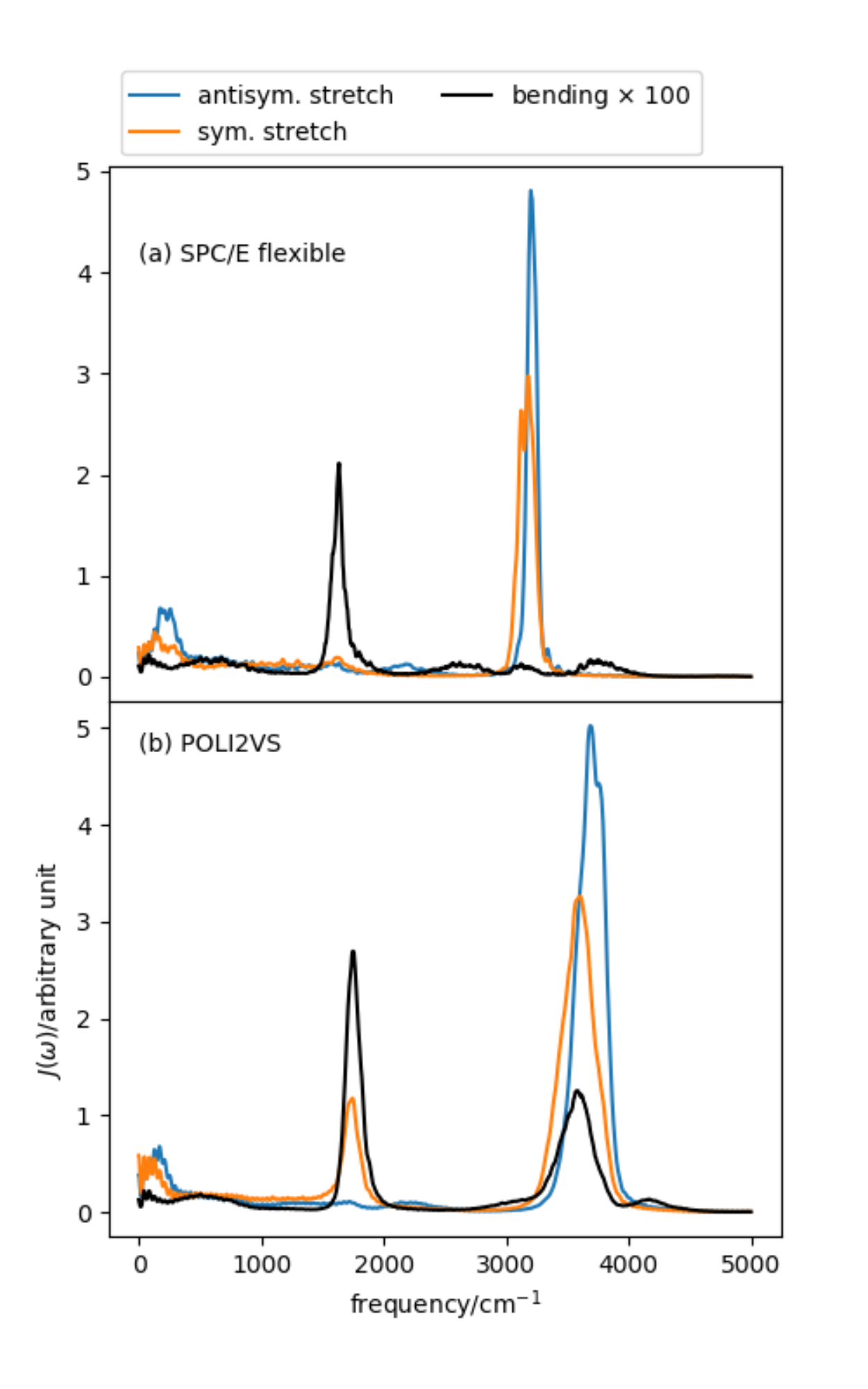}
\caption{The spectral distribution functions, $J_s (\omega)$, of the intramolecular water system plus bath model for the symmetric-stretch (orange curve), asymmetric-stretch (blue curve), and bend (black curve) modes
 obtained from the MD trajectory data with (a) the SPC/E flexible and (b) the POLI2VS force fields.}
\label{fig:Jomega}
\end{figure}

In Fig. \ref{fig:Jomega}, we display the results of the SDF
 for the symmetric-stretch (orange curve), asymmetric-stretch (blue curve),
 and bend (black curve) of H$_2$O molecules evaluated for (a) the SPC/E flexible and (b) POLI2VS cases.
It should be noted that the present results for each intramolecular mode with the LL+SL interaction are similar
 to those obtained in the case with the LL interaction only (not presented).
This indicates that the SL interaction is important in the system dynamics, while it plays a minor role for the bath modes. 

In each case, the intramolecular symmetric-stretch, asymmetric-stretch, and bend peaks are observed near 3100cm$^{-1}$, 3200cm$^{-1}$, and 1600cm$^{-1}$, respectively,
 while the intermolecular hydrogen bonded (HB) librational and HB translational peaks are observed in the range 800-400 cm$^{-1}$ and near 300 cm$^{-1}$, respectively.
In each intramolecular mode, we observe the peak at a frequency similar to that of its own mode,
 because each intramolecular mode can interact with the modes of the surrounding water molecules as an intermolecular interaction.

For the asymmetric-stretch mode (blue curve), the peak near 2200 cm$^{-1}$ arises from the coupling between the HB translation and the bend modes,
 whereas, for the bend mode (black curve), the peak near 3800 cm$^{-1}$ and 2600 cm$^{-1}$ arise from the coupling
 between the HB liberation and symmetric-stretch modes and the HB translation and bending modes, respectively.
These combination bands indicate the existence of energy transfer pathways among the modes:
 They are predicted on the basis of analysis of the energy transfer process.\cite{Yagasaki2011}
In the case of POLI2VS, we further observed combination bands between the bend and asymmetric-stretch modes and the symmetric- and asymmetric-stretch modes
 in the ranges 5000-5500 cm$^{-1}$ and 7000-7500 cm$^{-1}$, respectively. (See Appendix).
The assigned intermolecular modes and combination bands are considered in Table \ref{table:found_mode_bending}.
In the POLI2VS case, we found that the combination band peak of the HB transitional mode and bending mode is merged
 with the symmetric-stretch mode peak, and together they appear as a broadened peak.
Such features that describe the coupling between the inter- and intra-molecular vibrational modes have not been thoroughly explored
 on the basis of the infrared and Raman spectroscopic analysis, because they are optically inactive.
The present machine learning approach has the capability to estimate optically inactive combination bands
 that may be different for different force fields and different simulation approaches. 

The intermolecular mode frequencies obtained using POLI2VS are higher than those obtained using SPC/E,
 because POLI2VS was constructed for quantum simulations, while here we used it for classical simulations.
For this reason, the appearance of the combinational bands is also different in the POLI2VS case.
In order to obtain an accurate description, we must employ quantum trajectories.\cite{JJLiu18}

\begin{table}[htbp]
  \caption{Optimized potential parameters.}
  \label{table:optimized_param}
  \begin{tabular}{cccc}
    \hline
                      & & SPC/E flexible & POLI2VS   \\ \hline \hline
    & $k_t$ [kJ/mol/rad$^2$] & $382.168$    & {$364.963$} \\
    &  $k_r$ [kJ/mol/nm$^2$] & $311708 $    &    -      \\
    & $a$ [nm$^{-1}$]        &  -           & {$21.1293$} \\\hline
                                  & Bend. [rad$^{-1}$] & $-0.0485218$ & { $ 0.17656853$  } \\
    $V_\mathrm{SL}/V_\mathrm{LL}$ & Sym.  [nm$^{-1}$]  & $ 10.8475$   & { $ 18.41797858$ } \\
                                  & Asym. [nm$^{-1}$]  & $-1.73348$   & { $ -0.219222965$} \\\hline
              & $q_\mathrm{bend}q_\mathrm{sym}$  [kJ/mol/rad/nm] & $-217.080$ & { $-468.367$} \\
    $g_{ss'}$ & $q_\mathrm{bend}q_\mathrm{asym}$ [kJ/mol/rad/nm] & $-2.03171$ & { $ 0.06029$} \\
              & $q_\mathrm{sym}q_\mathrm{asym}$  [kJ/mol/nm$^2$] & $-103.004$ & { $-516.444$}  \\ \hline
                & $q_\mathrm{bend}^2q_\mathrm{sym}$  [kJ/mol/rad$^2$/nm] & $-212.825$ &{ $-3019.51$} \\
    $g_{sss'}$  & $q_\mathrm{bend}^2q_\mathrm{asym}$ [kJ/mol/rad$^2$/nm] & $ 101.093$ &{ $ 100.569$} \\
                & $q_\mathrm{sym}^2q_\mathrm{asym}$  [kJ/mol/nm$^3$]     & $ 250917$ & {$-4511.28$ }\\ \hline
                & $q_\mathrm{bend}q_\mathrm{sym}^2$  [kJ/mol/rad/nm$^2$] & $ 4123.39$ &{ $ 64048.4$} \\
    $g_{ss's'}$ & $q_\mathrm{bend}q_\mathrm{asym}^2$ [kJ/mol/rad/nm$^2$] & $-624.620$ &{ $-14091.7$} \\
                & $q_\mathrm{sym}q_\mathrm{asym}^2$  [kJ/mol/nm$^3$]     & $ 241830 $ &{ $ 345203 $} \\ \hline
  \end{tabular}
\end{table}

\begin{table}[htbp]
  \caption{Peak positions for the assigned intermolecular modes and convention bands observed in the bend mode.
           Results for the intramolecular bending and stretching modes are not presented.}
  \label{table:found_mode_bending}
  \begin{tabular}{ccc}
    \hline
    Mode               & SPC/E Flexible & POLI2VS \\ \hline \hline
    HB stretching mode & 85 cm$^{-1}$   &{ 69 }cm$^{-1}$ \\
    Rotation           & 557 cm$^{-1}$  &{ 509 }cm$^{-1}$  \\
    HB trans. + bend.  & 2623 cm$^{-1}$ & - \\
    HB lib. + sym.     & 3759 cm$^{-1}$ &{ 4170 }cm$^{-1}$ \\ \hline
  \end{tabular}
\end{table}

In the present system-bath model, it is not necessary that the frequency of the system mode be the same as the frequency of the molecular mode in the MD simulations,
 because the system-bath interaction modifies the central frequency of the system oscillator, as has been found in investigations
 employing Brownian motion theory\cite{Weiss12} and optical spectroscopic theory with the LL and SL interactions.\cite{Okumura-PRE-1997-56, dephase04,dephase05,dephase06,dephase07,Sakurai-JPCA-2011-115}
Although the peak heights in the evaluated spectral distribution may depend on the conditions of the optimization, for example, on the value of $\Delta t$ in the HOB simulations, the overall features of the results are insensitive to such details.
Verifying the validity of the system Hamiltonian and the system-bath interaction must be done in a case-by-case manner by, for example, evaluating the reaction rates and vibrational spectrum in any given application of interest

\section{6. CONCLUSION}
\label{sec:conclusion}

We studied a system-bath model to simulate the dynamics of molecular liquids characterized by molecular trajectories obtained from MD simulations.
We demonstrated that the machine learning approach is a versatile tool for estimating the model parameters
 for the system potential, mode-mode interactions, the system-bath coupling, and, most importantly, the SDFs of the bath.
An optimized system-bath Hamiltonian evaluated with the present approach allows us to carry out numerically expensive simulations that are not feasible in the case of the full quantum MD simulations,
 for example for calculations of 2D infrared\cite{Sakurai-JPCA-2011-115,Ishizaki-JCP-2006-125,2DIR-T01,2DIR-T02,Tokmakov2DIRH2O,Skinner2DIRH2O,Dwayne2DIRH2O}
 and 2D infrared-Raman spectroscopies,\cite{2DIRr,Bonn2DIRRaman,Radzewicz2DIRRaman} where the quantum dynamics of intramolecular modes play a significant role.

The obtained SDFs can also be used for characterization of the dynamics that play an important role in vibrational spectroscopy and energy translation
by taking account of the various combination bands that cannot be explored by spectroscopic means because they are optically inactive.
In particular, we discovered the existence of intermolecular interactions between the stretching modes of the system and the bath modes that consist of the stretching modes of the surrounding water molecules.

The difference between the model parameters and SDFs for different MD force fields indicates that we can use such information to verify the description of the MD force fields in a critical manner.
Moreover, if we use trajectories obtained from quantum dynamical simulations instead of classical simulations,
 such as {\it ab initio} MD simulations\cite{PaesaniIuchiVoth2007,Paesani2017,PIMD,AIMD01} or quantum MD simulations using a force field designed for quantum simulations\cite{JJLiu18,MLMD01, MLMD02, MLMD03, MLMD04},
 we should be able to obtain a system-bath Hamiltonian more suitable for quantum dynamical simulations. 

Here, we employed the HO and MO Hamiltonians with an anharmonic mode-mode interaction as a trial system.
Instead of these, we could also consider a system potential with the same form as that in the force field model used in the MD simulations.
Although such a system-bath Hamiltonian is much more difficult to study on the basis of open quantum dynamics theories,
 we found that such a model can describe the MD results better than the simple HO and MO models (results not presented).
In order to model a complex system more accurately, we should include physical observables,
 such as the optical dipole and polarization, in order to obtain the optimized model 
Methodologies and model functions developed in various machine learning approches, for example an approach developed for the physics-informed neural network (PINN),\cite{PINN, PINN2, DeepXDE} may be useful. 

The key feature of the present study is that it treats a system-bath model, in which the primary molecular dynamics are described by a system Hamiltonian,
 while the other degrees of freedom are described by a harmonic heat bath that is characterized by SDFs.
This approach can be applied not only to systems described by molecular coordinates,
 such as chemical reaction systems\cite{Wolynes81,Miller1989,HanggiRMP90, BOmodel03} and vibrational systems\cite{Mukamel95,Khun95,TaniIshi09},
 but also systems described by electronic states, such as systems exhibiting non-adiabatic transitions\cite{Onuchic85,Wolynes87ET, Ikeda18, Ikeda19},
 quantum devices\cite{MasonHess89, SakuraiNJP14}, ratchet rectification\cite{Hanggi97,Hanggi09, KatoJPCB13},
 and exciton-transfer systems\cite{Ishizaki09,Schuten12, Renger06, KramerAspu13, KramerAspu14, Coker2016}
 in situation that we have microscopic data describing not only molecular coordinates but also electronic states.\cite{Pavlo}
As a future investigation, we plan to extend the present study in such a direction.

\begin{acknowledgement}
The authors would like to thank Dr. Taisuke Hasegawa for providing the source code for the POLI2VS potential model.
YT is thankful to professor Shinji Saito for helpful discussions concerning the assignment of the combination bands appearing in the spectral distribution functions.
Financial support from HPC Systems Inc. is acknowledged.
\end{acknowledgement}

\appendix
\section{Appendix: Overtone of OH stretching in the POLI2VS model}

\begin{figure}[htbp]
\includegraphics[bb=0 0 403 310, width=8cm]{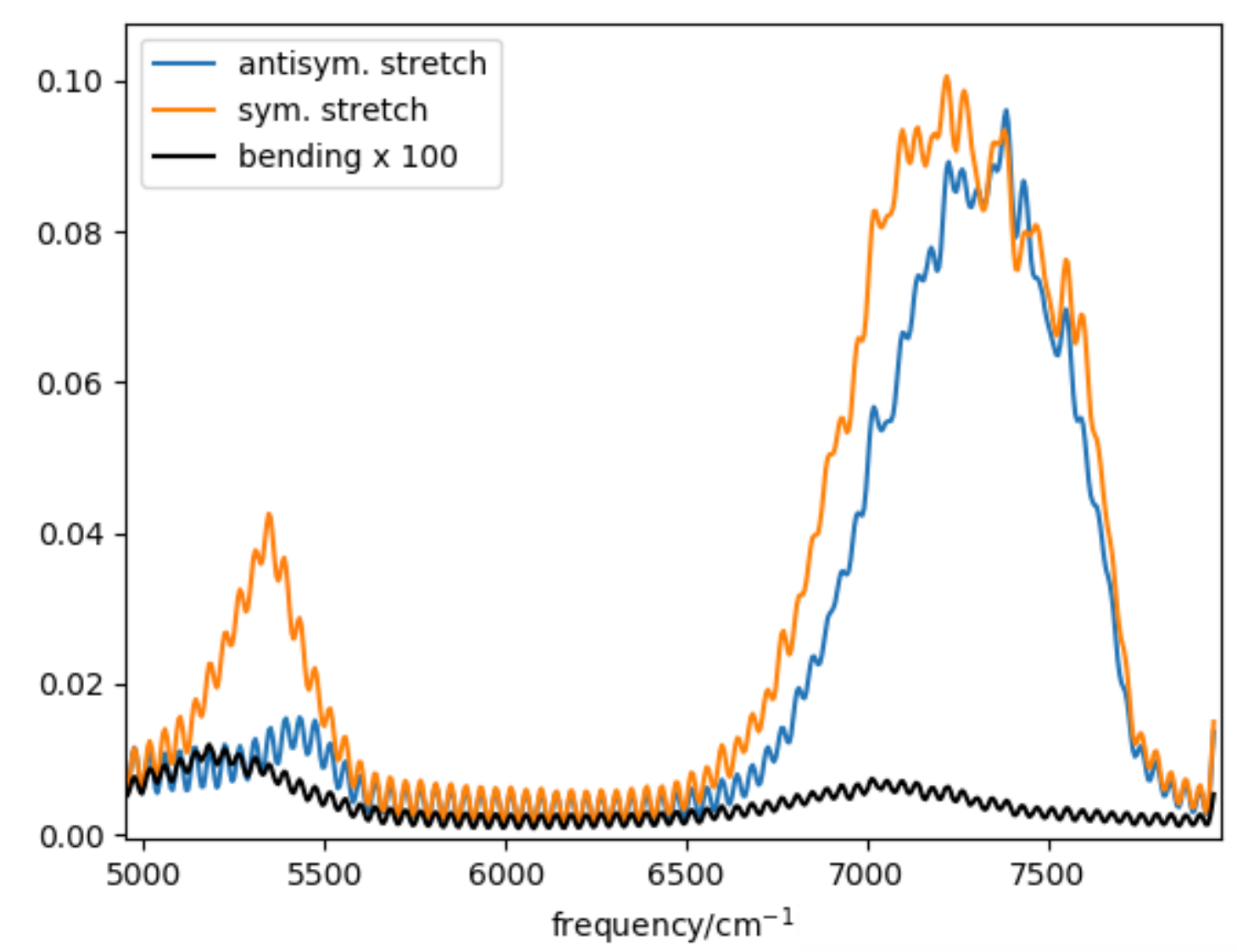}
\caption{Overtone line shapes of the SDFs obtained from the POLI2VS trajectories.}
\label{fig:Jomega_overtone}
\end{figure}
Figure \ref{fig:Jomega_overtone} illustrates the high-frequency overtone area of $J_s(\omega)$ obtained from POLI2VS.
(The low-frequency part is presented in Fig. 2(b).)
In the SPC/E flexible case, we could not observe these peaks.
In the symmetric and asymmetric modes, small peaks appear as overtones of the OH stretching mode.
The peak intensities of these overtones become larger if we employ the harmonic potential for the OH bonds instead of the Morse potential.
This indicates that the anharmonicity of the MD potential, which cannot be taken into account by the system potential,is described by the overtone peak of $J(\omega)$.

\bibliography{refs}

\end{document}